\documentclass[a4paper,12pt]{article}

\usepackage{graphicx} 
\usepackage{amsmath,amssymb}
\usepackage{ascmac}
\usepackage{cancel}
\usepackage{bm}
\usepackage{color}
\usepackage{braket}
\usepackage[nocompress]{cite}
\usepackage[colorlinks=true,linkcolor=blue,citecolor=blue]{hyperref}
\usepackage{arydshln}
\usepackage{multirow}
\usepackage{comment}
\usepackage{CJKutf8}
\usepackage{tikz}
\usepackage{tcolorbox}
\tcbuselibrary{theorems,skins}

\renewcommand{\rm}[1]{\textrm{#1}}
\newcommand{\ev}[1]{\Braket{#1} }
\newcommand{\V}[1]{{\bm{#1}}}	
\newcommand{\p}{\V{p}}

\renewcommand{\k}{\V{k}}
\newcommand{\x}{\V{x}}
\renewcommand{\v}{\V{v}}

\newcommand{\deltad}{\delta_{\rm D}{}}

\newcommand{\tdelta}{\tilde \delta}
\newcommand{\ttheta}{\tilde \theta}
\newcommand{\fh}{{\mathcal F H}}
\newcommand{\calf}{{\mathcal F}}

\makeatletter

\@addtoreset{equation}{section}
\makeatother

\begin{document}

\begin{titlepage}

\begin{center}

\vskip .45in

\bigskip\bigskip
{\Large \bf An Effective Theory for Biased Tracers 
\\[3mm]
via the Boltzmann-Equation Approach
}

\vskip .65in

{\large 
Tomohiro Fujita$^{1,2}$, 
Tomo Takahashi$^{3}$, and 
Sora Yamashita$^{4}$
\vspace{2mm} \\
}
\vskip 0.2in

{\it 
$^{1}$Department of Physics, Ochanomizu University, Bunkyo, Tokyo 112-8610, Japan
\vspace{2mm}\\
$^{2}$Kavli Institute for the Physics and Mathematics of the Universe (Kavli IPMU), WPI, UTIAS, The University of Tokyo, Kashiwa, Chiba 277-8568, Japan
\vspace{2mm}\\
$^{3}$Department of Physics, Saga University, Saga 840-8502, Japan
\vspace{2mm}\\
$^{4}$Graduate School of Science and Engineering, Saga University, Saga 840-8502, Japan
}

\end{center}
\vskip .5in

\begin{abstract}
We develop an effective theory for biased tracers formulated at the level of the Boltzmann equation, providing a unified description of density and velocity bias. We introduce a general effective collision term in the tracer Boltzmann equation to encode tracer dynamics that are intrinsically different from those of dark matter. This collision operator leads to modified continuity and Euler equations, with source terms reflecting the collision-term physics. At linear order, this framework predicts time- and scale-dependent bias parameters in a self-consistent manner, encompassing peak bias as a special case while clarifying how velocity bias and higher-derivative effects arise. Applying the resulting bias model to redshift-space distortions, 
we compare the power spectrum of biased tracers with that obtained in the Effective Field Theory of Large-Scale Structure.

\end{abstract}

\end{titlepage}

\section{Introduction \label{section:introduction}}

Large-scale structure (LSS) serves as one of the most powerful probes of cosmology \cite{Peebles:1980yev,Bernardeau:2001qr}, providing a detailed record of how matter in the Universe has evolved from its early, nearly homogeneous state to the rich structure observed today. This evolution is primarily driven by dark matter 
(DM), 
which constitutes about 80\% of the matter content. The growth of dark-matter fluctuations has been extensively studied through theoretical modeling and simulations, and their statistical properties are now understood in considerable detail.
However, we cannot directly observe the dark matter distribution. Except through gravitational lensing, observations capture only the luminous baryonic component such as galaxies which follow but do not perfectly trace the underlying matter distribution. Such observable objects are referred to as biased tracers, and understanding their behavior is crucial for extracting reliable cosmological information from LSS observations (see \cite{Desjacques:2016bnm} for a review).

Biased tracers that we observe, such as the DM halos, galaxies, and clusters, not only trace the DM evolution but also form, merge, and are disrupted. Although we can predict the DM distribution and its evolution, they are different from those of biased tracers. To describe biased tracers, we should construct an appropriate bias model that connects DM and tracers.
To find such a bias model, one effective approach is to work with two fluid equations, the continuity and Euler equations, obtained from the Boltzmann equation by applying the non-relativistic limit. Since the number of biased tracers is not conserved (because of formation, mergers, and disruption), their fluid equations should be modified so that number conservation is violated by suitable source terms. 
Moreover, the Euler equation could also require modification to include effective contributions not captured by the direct large-scale gravitational interaction, for example those induced by baryonic processes.
Such modifications by additional source terms have been discussed in \cite{Fujita:2020xtd,Kehagias:2013rpa,Chan:2012jj}.

The density fluctuation of a biased tracer, $\delta_g$, is often modeled as a functional of the matter overdensity, $\delta$, namely $\delta_g = b[\delta]$. In practice, $b[\delta]$ is usually specified by phenomenological bias models such as local bias \cite{Fry:1992vr}, nonlocal bias \cite{McDonald:2009dh,Chan:2012jj}, Lagrangian bias \cite{Bardeen:1985tr,Mo:1995cs,Matsubara:2008wx}, and the bias in the Effective Field Theory of Large-Scale Structure (EFTofLSS)~\cite{Senatore:2014eva}.
In addition to the density bias, velocity bias has also been discussed, e.g., in the peak bias model \cite{Desjacques:2009kt,Baldauf:2014fza,Elia:2011ds,Matsubara:2019tyb}. However, large-scale velocity bias is usually assumed to be absent simply as a consequence of the equivalence principle~\cite{Creminelli:2013nua,Desjacques:2016bnm,Chan:2012jj}.

As explained above, the fluid equations for dark matter are obtained from the Boltzmann equation. Motivated by this, in this work we do not model bias phenomenologically at the level of the density and velocity fields, but instead develop an effective description directly at the level of the Boltzmann equation. We refer to this framework as the {\it Boltzmann Equation Approach} (BEA). Within the BEA, we derive effective fluid equations for biased tracers, which contain source terms, by incorporating collision terms in the Boltzmann equation that characterize the evolution of the distribution function for biased tracers. In this way, the source terms that appear in the tracer fluid equations are fixed once the collision terms are specified, and the same underlying dynamics simultaneously determine both the density and velocity fields of the tracers. As a result, the density bias and velocity bias are obtained in a self-consistent manner from a single dynamical framework, which is one of the main advantages of the BEA. The goal of this work is to use the BEA to construct effective fluid equations for biased tracers and to compute the resulting observables such as the power spectrum, and compare our results with those obtained in existing approaches, including the EFTofLSS~\cite{Perko:2016puo,Ivanov:2019pdj}.

This paper is organized as follows. In Section~\ref{section:review}, we review how the fluid equations are derived from the Boltzmann equation. In Section~\ref{section:biased-boltzmann}, we construct a Boltzmann equation that determines the tracer’s distribution, assuming a collision term to derive the fluid equations. Then we predict the bias evolution in the linear regime, checking its consistency with the peak bias \cite{Bardeen:1985tr}. In Section~\ref{section:power-spectrum}, 
we apply our formalism to predict the power spectrum in redshift space, and compare the results with those in the EFTofLSS.
The final section is devoted to conclusion of this paper.

\section{The fluid equations from Boltzmann equation \label{section:review}}

First, we briefly review how to obtain the DM fluid equations from the Boltzmann equation \cite{Dodelson:2020bqr,Carrasco:2012cv,Hertzberg:2012qn,Baumann:2010tm}. The relativistic collisionless Boltzmann equation
is written as follows:
\begin{align}
  \left(
    P^\mu \frac{\partial}{\partial x^\mu} 
    - \Gamma^\mu{}_{\alpha\beta}P^\alpha P^\beta\frac\partial{\partial P^\mu}
  \right)f=0,
  \label{boltzmann-rela}
\end{align} 
where $x^\mu,~P^\mu,~\Gamma^\mu{}_{\alpha\beta},~f$ are the 4-coordinate, 4-momentum, Christoffel symbol, and the distribution function, respectively.
We adopt the Newtonian gauge for the perturbed Friedmann--Lema\^{\i}tre--Robertson--Walker metric and then, its invariant line element is given by
$ds^2  = -(1+2\Phi)dt^2 + a^2 (1-2\Psi) d\x^2$ where $a(t)$ is the scale factor normalized to unity at the present time and $\Phi$ and $\Psi$ are the metric perturbations, $|\Phi|, |\Psi|\ll 1$. 
Here we assume that the anisotropic stress of matter fluid can be neglected so that $\Phi=\Psi$, and the DM momentum $P^\mu = (P^0 \,, {\bm P})$ satisfies
\begin{align}
   |\V{P}| \ll m,\qquad P^0\simeq m\,.
\end{align}
In the following, we work in  the Newtonian limit where $ |\dot \Phi| \ll |\nabla\Phi|$ holds,
the Boltzmann equation is approximated as
\begin{align}
    \frac{Df}{Dt}:=\dot f + \frac{1}{ma^2}(\p\cdot\nabla)f - m\nabla\Phi\cdot \nabla_pf=0\,,
    \label{vlasov-dm}
\end{align}
with $\p:= a^2 \V P$ being the canonical comoving momentum and $\nabla_p := \left(\partial_{p^1},\partial_{p^2},\partial_{p^3}\right)$.
Here, the gravitational potential $\Phi$ satisfies the Poisson equation:
\begin{align}
    \Delta \Phi = 4\pi G a^2 \bar\rho \delta\,.
    \label{poisson-eq}
\end{align}
$\bar{\rho}$ and $\delta$ are defined soon below.
By integrating $f(t,\p,\x)$ over $\p$, we can obtain the number density (or the energy density) and velocity field  as the 0-th and first moments:
\begin{align}
    & \int_\p := \int d^3p\,, \\
    &\int_\p f(t,\p,\x) 
      = a^3n(t,\x) = a^3\frac{\rho(t,\x)}{m} = a^3\frac{\bar\rho(t) + \bar\rho\delta(t,\x)}{m}\,, 
    \label{zero-moment}
    \\
    &\int_\p p^i f(t,\p,\x) 
      = a^4\rho(t,\x)v^i(t,\x)\,, 
    \label{one-moment}
\end{align}
where $\bar \rho(t)$ is the spatial average of the energy density and $\delta(t,\bm x)$ is the density contrast, $\rho(t,\bm x)=\bar \rho(t)+\bar\rho\delta(t,\bm x)$.
$v^i$ is the velocity field defined by Eq.~\eqref{one-moment}.
The second moment can be approximated as follows, especially 
for pressureless perfect fluid \cite{Baumann:2010tm,Weinberg:1972kfs}:
\begin{align}
    \int_\p p^ip^j f(t,\p,\x) 
      \simeq ma^5\rho v^i v^j.
\end{align}
Using the above relations, we obtain
\begin{align}
    &\begin{aligned}
      0 = \int_\p \frac{Df}{Dt}
      = \frac{a^3}{m}
      \left[
        \dot \rho + 3H\rho + \frac1a\nabla\cdot  \rho\v  
      \right],
    \end{aligned}
      \\
    &\begin{aligned}[b]
      0 
      = \int_\p p^i \frac{Df}{Dt}
      &  
      =  \frac{d}{dt}\left(a^4\rho v^i \right)
        + a^3 \left[
          \partial_j(\rho v^j)\, v^i + \rho v^j\partial_j\,v^i 
        \right]
        + a^3\rho \partial_i\Phi 
      \\
      & = a^4\left[
          \left(H\rho - \frac1a \nabla\cdot  \rho \v \right)v^i
          + \rho  \dot v ^i
        \right] 
        + a^3 \left[
          \partial_j(\rho v^j)\, v^i + \rho v^j\partial_j\,v^i 
        \right]
        + a^3\rho \partial_i\Phi \,,
    \end{aligned}
\end{align}
where $H:=\dot a/a$ is the Hubble parameter. The first equation implies $\dot{\bar\rho} + 3H\bar\rho=0$ and hence $\bar\rho \propto a^{-3}$ at the background level. Expanding the equations with $\rho = \bar\rho +\bar\rho \delta$, 
we obtain the fluid equations in the expanding Universe as
\begin{align}
    &\dot \delta  + \frac1a \nabla\cdot\left[(1+\delta)\v\right]=0\,,
    \label{continu eq}
    \\
    &\dot \v + H\v + \frac1a (\v\cdot\nabla)\v +\frac1a\nabla\Phi=0\,.
    \label{Euler eq}
\end{align}
Moreover, taking the linear limit and combining Eqs.~\eqref{continu eq} and \eqref{Euler eq}, one obtains
\begin{align}
  &\ddot \delta + 2H \dot\delta -\frac32H^2\Omega_m\delta=0 \,,
  \label{second-order-delta-eq}
\end{align}
where Eq.~\eqref{poisson-eq} has been used, and the matter density parameter is defined as
\begin{align}
  \Omega_m(t):= \frac{8\pi G \bar\rho}{3H^2}.
\end{align}
Eq.~\eqref{second-order-delta-eq} yields a growing and a decaying mode, and the latter is negligible. The growing mode can be written as
\begin{align}
  \delta(t,\x) = D(t)\delta(t_\rm{ini},\x),
  \label{def-growth-factor}
\end{align}
where \(D(t)\) is the growth factor. The corresponding growth rate is defined as
\begin{align}
\label{eq:F}
  \mathcal F := \frac{\dot D}{HD}.
\end{align}

Before discussing the Boltzmann equation for biased tracers, we note that the phase-space distribution function can also be decomposed into its background and perturbation parts as
\begin{align}
    &f(t,\x,\p) = \bar f(t,|\p|) + \delta f(t,\p,\x)\,,
\end{align}
where $\bar f$ denotes the spatial average. 
The zeroth and first moments of $\bar{f}$ and $\delta f$ are given as follows:
\begin{align}
    &\int_\p \bar f = \frac{a^3\bar\rho}{m},
      \qquad 
      \int_\p \p\bar f =0,
      \qquad
      \int_\p \delta f =\frac{a^3\bar\rho \delta}{m},
      \qquad
      \int_\p \p\delta f =a^4\rho \v\,,
\end{align}
where we use the fact that $\bar f(t,|\p|)$ 
only depends on the magnitude of $\p$.

\section{Boltzmann equation for the biased tracer \label{section:biased-boltzmann}}

In Sec.~\ref{section:review}, we derived the fluid equations for dark matter from its collisionless Boltzmann equation.
For biased tracers, however, 
the same equations cannot be adopted,
because their evolution generally includes processes such as formation and merging, and thus their number is not conserved.
These effects imply that the effective continuity and Euler equations for tracers should be modified from those of dark matter, and that both density and velocity biases can emerge.
To account for these differences in a systematic way, we now extend the Boltzmann equation by introducing an effective collision term that captures the distinct dynamics of biased tracers.

Since biased tracers, by definition, follow the underlying dark-matter distribution, it is reasonable to assume that their statistical behavior can be characterized by the dark-matter phase-space distribution itself.
From this perspective, we represent the collision term in the tracer's Boltzmann equation as an effective functional of the dark-matter distribution.
Since the tracer sector is a system subject to formation/merger/disruption/tidal stripping/selection, the collision term is modeled statistically by an effective gain--loss (transition-kernel) description, such that small distance phenomena are considered to be integrated out. In light of this general idea, we systematically approximate the collision term with long wavelength and small momentum transfer expansion to leading order around   the homogeneous background as follows:
\begin{align}
    \frac{Df_g}{Dt}= {}
        &{}
      \fh C_1\delta f 
      + C_2 \dot f
      + \frac{C_3}\fh \Delta f 
      + ma^2 C_4 \nabla\cdot \nabla_p f 
      + (\fh)^3C_5 \Delta_p f\,,
    \label{vlasov-gal}
\end{align}
where $\Delta_p := \nabla_p^2$, $f_g$ denotes the tracer distribution function, and $C_1,  \dots,  C_5$ are dimensionless time-dependent coefficients that capture unresolved small-scale physics. The factors such as $ma$ and $\fh$ are introduced to make the coefficients dimensionless.

The first term is introduced to represent a linear response against the
disturbance due to dark matter perturbations. Although it appears to be the so-called relaxation term where the system is close to equilibrium (e.g., \cite{Jaiswal:2013npa}), in the present context,
this term should not be interpreted as a relaxation term toward an
equilibrium state.  
Instead, it parametrizes the leading local-in-time
response of the tracer distribution to the perturbation of the
underlying dark-matter distribution.
The sign of this response is not fixed a priori: it may either reduce or
enhance the difference between \(f_g\) and \(f\). 

The second term describes a memory effect, or non-local response in time, which arises from 
the expansion in a small time scale $\Delta t$, corresponding to that of the dynamics of the tracer much shorter than the background cosmic expansion,  $f(t-\Delta t)$, yielding $\dot{f}$.

The third term stems from long wavelength 
or spatial derivative expansion, in which, in the Fourier space, $k R_\ast \ll 1$ with $R_\ast$ being the scale under which the tracer system is affected by the small scale phenomena 
such as formation and merger. This expansion can be schematically described as
\begin{align}
    \mathcal C[f](\x)
    =
    \int d^3\x' \,\,
    K(|\x'|) 
    \left[
     f(\x) + x'^i \nabla_i f(\x) 
     + \frac12 x'^i x'^j \nabla_i \nabla_j f(\x) + \dots
    \right] \,,
  \end{align}
where $K(|\x|)$ is an isotropic kernel. 
The first term in the above equation can be absorbed in the first term in RHS of Eq.~\eqref{vlasov-gal}, while the linear in $x'^i$ vanishes after integration, leaving only the term $ x'^i x'^j \nabla_i \nabla_j f$.

The fourth and fifth terms are motivated by the expansion with respect to momentum derivative where we assume that the momentum transfer would be small.  
This can be written as
 \begin{align}
    \mathcal C[f](\p)
    =
    \int d^3(\Delta\p)
    \left[
      w(\p-\Delta\p,\Delta\p)f(\p-\Delta\p)
      - w(\p,\Delta\p)f(\p)
    \right]
  \end{align}
where $w(\p,\Delta\p)$ is the transition rate for $\p\to \p+\Delta\p$ with $\Delta \p$ being the momentum transfer in the process involved in the tracer dynamics on small scales. 
Introducing a typical momentum scale (e.g.\ set by the virial velocity), we assume $|\Delta \p|$ is small compared to this scale, so that the integrand can be Taylor-expanded for $|\p| \gg |\Delta \p|$. After coarse-graining the tracer dynamics, we assume that the effective momentum transfer is small, and that the distribution is isotropic in momentum space. Therefore the term with $\nabla_p f$ disappears and the leading term provides a dominant contribution. We also include the operator $\nabla\cdot\nabla_p$ in the framework where second-order spatial and momentum derivatives are retained, 
which makes the equation more general.
Indeed, there are well-known examples of this type such as Fokker-Planck equation \cite{DiPerna1988} and the one that describes cosmic-ray distributions \cite{Amato:2017dbs}. 

With these considerations in mind, we write down the collision term,  which is assumed to be expressed up to the second-order spatial and momentum derivatives, and the evolution of tracers should depend not only on the local coordinate and momentum, but also on their non-local environment,  
as RHS of Eq.~\eqref{vlasov-gal}.
Since the formation and motion of tracers are influenced by the past history of the underlying dark-matter field, their effective interaction must carry information from neighboring regions and earlier times, which can be approximately captured through a derivative expansion\footnote{
The spirit of this argument parallels the treatment of non-locality in time and in space often discussed within the framework of the EFTofLSS~\cite{Senatore:2014eva}.
}. These operators may account for the Fingers of God effect in redshift space 
because of their diffusive nature.

To write down the continuity and Euler equations for the biased tracers, we need to calculate the zeroth and first moments of the collision terms. The zeroth moments are given by
\begin{align}
    &\int_\p \dot { f} 
      = \frac{d}{dt} \frac{\rho a^3}{m}
      = \frac{\bar\rho a^3}{m}\dot \delta
      = \int_\p \dot {\delta f}\,,
      \qquad (\because ~\bar\rho a^3=\textrm{const})
      \\[8pt]
    &\int_\p \Delta { f} 
      = \Delta \frac{\rho a^3}{m}
      = \frac{\bar\rho a^3}{m}\Delta \delta
      = \int_\p \Delta {\delta f}\,,
      \\[8pt]
    &\int_\p \nabla\cdot \nabla_p { f} 
      =
      \int_\p \nabla\cdot \nabla_p {\delta f}
      = 0\,,
      \\[8pt]
    &\int_\p \Delta_p { f}
      =
      \int_\p \Delta_p {\delta f}
      =0\,.
\end{align}
The first moments can be calculated as
\begin{align}
    &\int_\p \p \dot { f} 
      = \int_\p \p \dot {\delta f} 
      = \frac{d}{dt} (\rho\v a^4)\,,
      \\[8pt]
    &\int_\p \p \Delta { f} 
      = \Delta \int_\p \p \delta { f}
      = \Delta (\rho\v a^4)\,,
      \\[8pt]
    &\int_\p \p\nabla\cdot \nabla_p { f} 
      =
      \int_\p \p \nabla\cdot \nabla_p {\delta f}
      = -\frac{\bar\rho a^3}{m}\nabla\delta\,,
      \\[8pt]
    &\int_\p \p\Delta_p { f}
      =
      \int_\p \p\Delta_p {\delta f}
      =0\,.
\end{align}
Therefore the continuity and Euler equations for the biased tracer are given by 
\begin{align}
    &\dot \rho_g + 3H\rho_g + \frac1a \nabla\cdot (\rho_g \v_g) 
    = 
      \fh C_1\bar\rho\delta 
      + C_2\bar\rho\dot \delta 
      + \frac{C_3}\fh\bar\rho\Delta\delta 
      + \frac{m^2 C_4}{a}\cdot0
      + m\left(\frac{\fh}{a}\right)^3 C_5\cdot0\,, 
    \label{continuity-eff}
    \\
    &\dot \v_g + H\v_g +\frac1a(\v_g\cdot\nabla)\v_g 
    +\frac1a\nabla\Phi
    \nonumber
    \\
    &
    =\frac{\rho}{\rho_g}
    \left[
      \fh C_1\v 
      + C_2 \left(
        \dot \v 
        + H\v 
        - \frac{\nabla\cdot(\rho\v)}{a\rho}\v 
      \right)
      + \frac{C_3}\fh \frac{\Delta(\rho\v)}{\rho}
      - a C_4\frac{\nabla\delta}{1+\delta}
      + \frac{(\fh)^3C_5\cdot 0}{a^4\rho}
    \right]\,,
    \label{eff-euler}
\end{align}
where we have explicitly written the terms that vanish after integration over  phase space.
We can express the linearized fluid equations of \eqref{continuity-eff} and \eqref{eff-euler} simply as follows:
\begin{align}
    &\dot\delta_g + \frac1a{\nabla\cdot \v_g} 
      = 
        \fh C_1\delta 
        - \frac1a C_2\nabla\cdot \v 
        + \frac{C_3}{\fh}\Delta\delta
        ,
    \label{continuity-gal-cosmic-time}
    \\
    &\dot\v_g + H\v_g + \frac1a{\nabla\Phi} 
      = 
      \fh C_1\v 
      - \frac1aC_2 \nabla\Phi 
      + \frac{C_3}{\fh}\Delta\v  
      - aC_4\nabla\delta
      ,
      \label{euler-gal-cosmic-time}
\end{align}
where we have applied the linearized DM fluid equations for $C_2$ term:
\begin{align}
    \dot\delta + \frac1a\nabla\cdot \v=0,
    \qquad
    \dot\v +H \v = -\frac1a\nabla\Phi.
    \label{linear-fluid-dm}
\end{align}
In the linear-order calculation above, we approximate $\rho/\rho_g \simeq \bar{\rho}/\bar{\rho}_g$ and  
absorb the factor $\bar{\rho}/\bar{\rho}_g$ into the definition of the coefficients as
\begin{align}
  \frac{\rho}{\rho_g}C_i \simeq \frac{\bar{\rho}}{\bar{\rho}_g}C_i \;\to\; C_i \qquad (i = 1,\dots,5)\,.
\end{align}
Note that the same symbol $C_i$ is used for the rescaled coefficients.
By introducing new quantities $\theta$ and $\theta_g$ defined as follows, 
\begin{align}
    \theta := -\frac{\nabla\cdot\v}{a\fh},
    \qquad
    \theta_g := -\frac{\nabla\cdot\v_g}{a\fh},
    \label{def-theta}
\end{align}
we obtain simple fluid equations in terms of Hubble parameter $H$, growth rate $\mathcal F$ defined in Eq.~\eqref{eq:F} and the density parameter $\Omega_m:= 8\pi G \bar\rho/(3H^2) $:
\begin{align}
    &\frac{1}{\fh}\dot\delta_g - \theta_g 
    = 
    C_1{\delta} + C_2\theta + \frac{C_3}{(\fh)^2}{\Delta\delta}\,,
    \label{fluid-delta}
    \\
    &\frac{1}{\fh}\dot\theta_g + \left(\frac32\frac{\Omega_m}{\calf^2}-1\right)\theta_g -\frac32\frac{\Omega_m}{\calf^2}\delta 
    =
    C_1{\theta} +  \frac32C_2\frac{\Omega_m}{\calf^2}\delta +
    \frac{C_3}{(\fh)^2} \Delta{\theta} 
    +\frac{C_4}{(\fh)^2}{\Delta\delta}\,,
    \label{fluid-theta}
\end{align}
where we used some relations to obtain the above equations:
\begin{align}
  & \ddot D + 2H\dot D - \frac32 H^2\Omega_mD=0 \,,
  \qquad \textrm{(obtained from \eqref{second-order-delta-eq} and \eqref{def-growth-factor})}
  \\
  &\partial_t ({\mathcal  F} H) 
    = - 2\mathcal FH^2 + \frac32H^2\Omega_m - (\fh)^2 \,,
  \\
  &\partial_t(\nabla\cdot \v_g)
    =
  -\partial_t(a\fh \theta_g)
    =
    -a\fh\left[
      \dot\theta_g - \fh\theta_g 
      \left(
        \frac {1}{\calf} - \frac32 \frac{\Omega_m}{\calf^2} + 1
      \right)
    \right] \,.
\end{align}
By changing the time variable to the scale factor,
the tracer’s fluid equations take the following form:
\begin{align}
    &\frac{a}{\mathcal F}\partial_a\delta_g - \theta_g
    = 
    C_1{\delta} + C_2\theta + \frac{C_3}{(\fh)^2}{\Delta\delta}\,,
    \label{eff-fluid-delta-a}
    \\
    &\frac{a}{\mathcal F}\partial_a\theta_g + \left(\frac32\frac{\Omega_m}{\mathcal F^2}-1\right)\theta_g -\frac32\frac{\Omega_m}{\mathcal F^2}\delta 
    =
    C_1{\theta} +  \frac32C_2\frac{\Omega_m}{\mathcal F^2}\delta +
    \frac{C_3}{(\fh)^2} \Delta{\theta} 
    + \frac{C_4}{(\fh)^2}{\Delta\delta}\,.
    \label{eff-fluid-theta-a}
\end{align}
In the matter-dominated (MD) Universe ($\mathcal F^2 = \Omega_m = 1$), 
Eqs.~\eqref{eff-fluid-delta-a} and \eqref{eff-fluid-theta-a} are expressed 
in a more simple manner:
\begin{align}
    &a\partial_a\delta_g - \theta_g 
    = 
    C_1{\delta} + C_2\theta + \frac{C_3}{H^2}{\Delta\delta} ,
    \\
    &a\partial_a\theta_g + \frac12\theta_g -\frac32\delta 
    =
    C_1{\theta} +  \frac32C_2\delta +
    \frac{C_3}{H^2} \Delta{\theta} 
    + \frac{C_4}{H^2}{\Delta\delta}.
\end{align}

We now move to the Fourier space and write down the equations for $\tilde\delta(t,\bm k)$.
Since we work in the linear regime, 
the evolution of perturbations for each wavenumber $k$ mode can be treated separately.
We then introduce effective bias parameters that may include
a $k$-dependence, defined by
\begin{equation}
    b(a,k) = \frac{\tilde{\delta}_g(k)}{\tilde{\delta}(k)}\,,
    \qquad
    b_v(a, k) = \frac{\tilde{\theta}_g(k)}{\tilde{\theta}(k)}\,.
\end{equation}
The evolution equations for $b$ and $b_v$ are given by
\begin{align}
    &a\partial_ab + (b - b_v) = C_1 + C_2 -C_3 (k/H)^2\,,
    \label{bias-equation-Fourier-b}
    \\
    &a\partial_ab_v  + \frac32({b_v} -1) 
    =
    C_1 + \frac32 C_2 -C_3 (k/H)^2 - C_4 (k/H)^2\,,
    \label{bias-equation-Fourier-bv}
\end{align}
where we used $\tdelta(k)=\ttheta(k)$, which holds for DM fluid.
These linear 
differential 
equations admit analytic formal solutions.  
We first obtain a solution for $b_v$ from the second equation, and then by using it we also obtain $b$:
\begin{align}
    &b(a,k) = 
      1-2 A(k)a^{-3/2} + B(k)a^{-1} 
      + C_g + C_{gk}(k/H_0)^2\,, \label{bias-result-g}
      \\
    &b_v(a,k) = 
      1+ A(k) a^{-3/2}
      + C_v + C_{vk}(k/H_0)^2\,,\label{bias-result-v}
\end{align}
with
\begin{align}
    &C_g(a) :=  
      \int d\tilde a~
        \left[
          a^{-1}(3C_1 + 4C_2)
          -a^{-3/2} \tilde a^{1/2}(2C_1 + 3C_2)
        \right]\,,
    \\
    &C_{gk}(a) := 
       -\int d\tilde a~ \frac{H_0^2}{H^2(\tilde a)}
        \left[
           a^{-1}(3C_3 +2C_4)
           - a^{-3/2}\tilde a^{1/2} (2C_3+2C_4)
        \right]\,,
    \\
    &C_v(a) :=  
      a^{-3/2}\int d\tilde a~
        \tilde a^{1/2} \left[
          C_1 + \frac32C_2
        \right]\,,
    \\
    &C_{vk}(a) :=  
      -a^{-3/2}\int d\tilde a~\frac{H_0^2}{H^2(\tilde a)}
        \tilde a^{1/2} \left[
          C_3 + C_4
        \right]\,,
\end{align}
where $A(k)$ and $B(k)$ are integration constants, determined by the initial conditions of $b$ and $b_v$.  
\begin{figure}[t]
    \centering
    \includegraphics[width=0.49\linewidth]{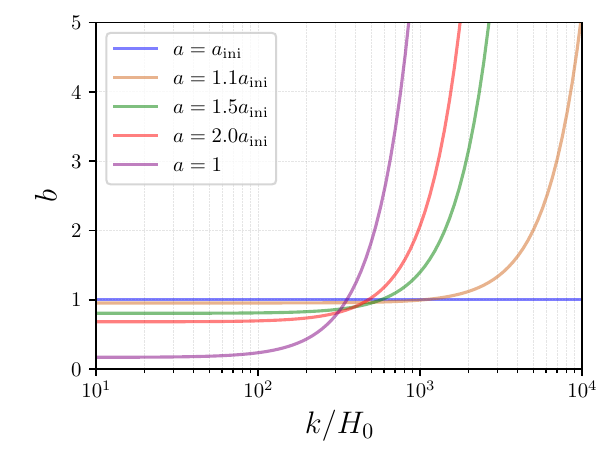}
    \includegraphics[width=0.49\linewidth]{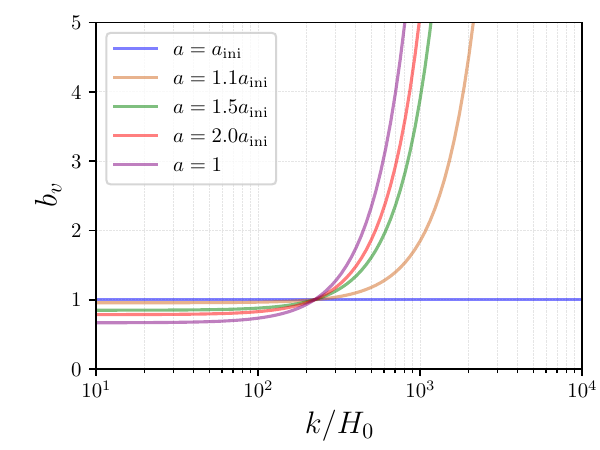}
    \caption{The time and $k$-dependence of the bias parameters $b(a,k)$ and $b_v(a,k)$ are shown in the left and right panels, respectively. The parameters are set to $C_1=-1/2,C_4H_0^2/H^2=-10^{-5},~C_2=C_3=0$ 
    for which the simple solutions in  Eqs.~\eqref{b sol} and \eqref{bv sol} are applicable. 
    We also set $A(k)$ and $B(k)$ so that $b=b_v=1$ at the initial time $a_\mathrm{ini}=10^{-3}$ for all $k$. 
    The initial $k$-dependent contributions of $A(k)$, $B(k)$, and $C_4$ cancel each other. As the system evolves, however, the $A(k)$ and $B(k)$ terms decay while the $C_4$ term persists, giving rise to the $k^2$ dependence seen at late times.
    }
    \label{fig:b-and-bv_eff-const}
\end{figure}
\begin{figure}[t]
    \centering
    \includegraphics[width=0.49\linewidth]{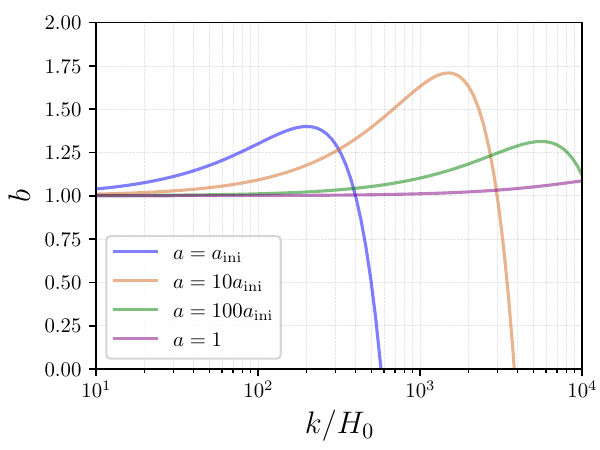}
    \includegraphics[width=0.49\linewidth]{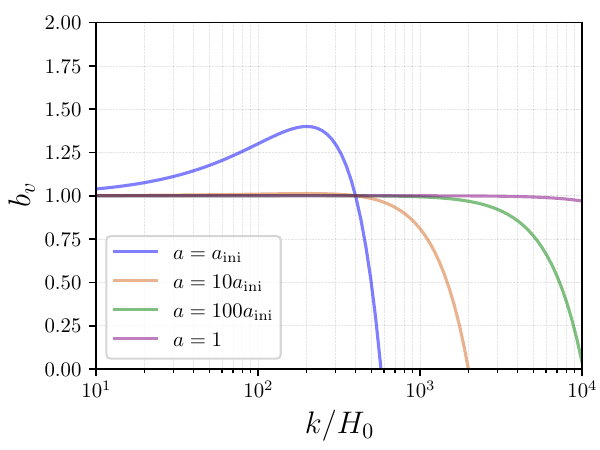}
    \caption{$b(a,k)$ and $b_v(a,k)$ are shown for the parameter set in Eqs.~\eqref{Fig2 para1} and \eqref{Fig2 para2}. Unlike Fig.~\ref{fig:b-and-bv_eff-const}, $b(a,k)$ and $b_v(a,k)$ exhibit non-trivial $k$-dependence at the initial time $a_{\mathrm{ini}}$, and the diffusion terms $C_{gk}$ and $C_{vk}$ are assumed to decay as $a^{-3/2}$. 
As seen in Eqs.~\eqref{b sol} and \eqref{bv sol}, the coefficients $A(k)$ and $B(k)$ always decay with time, and together with the diminishing diffusion terms, the initial $k$-dependence is eventually washed out. Consequently, the bias parameters gradually approach scale-independent behavior on large scales.
}
    \label{fig:b-and-bv_C-k-linear}
\end{figure}

In particular, if $C_1,~C_2,~C_3/H^2,~C_4/H^2$ are constant, the solutions take a more compact form:
\begin{align}
    &b(a,k) =
      1-2 A(k)a^{-3/2} + B(k)a^{-1}  +\frac53C_1 + 2C_2 -\frac53C_3 (k/H)^2 -\frac23 C_4(k/H)^2 ,
      \label{b sol}
      \\
   &b_v(a,k) =  
     1 + A(k) a^{-3/2} + \frac23\left[ C_1 + \frac32 C_2 -C_3 (k/H)^2-C_4(k/H)^2 \right]\,.
     \label{bv sol}
\end{align}
The $A(k)$ and $B(k)$ terms decay with time, whereas the $C_3$ and $C_4$ terms remain constant, irrespective of the initial condition.
Consequently, after a sufficiently long time,
the $k^2$-components become dominant on small scales. 
Fig.~\ref{fig:b-and-bv_eff-const} illustrates this behavior,
in which the initial conditions at $a=a_\mathrm{ini}=10^{-3}$ and the constant parameters are set as
\begin{align}
  b(a_\mathrm{ini},k)=b_v(a_\mathrm{ini},k)=1, 
    \label{bias-initial-condition-1}
  C_1=-1/2,~C_4H_0^2/H^2=-10^{-5},~C_2=C_3=0\,,
\end{align}
which correspond to $C_g=-5/6,~C_{gk}
= 1/15
\times 10^{-4}, C_v=-1/3,~C_{vk}
=1/15 \times 10^{-4}$.

Even when the assumption of constant coefficients is relaxed, this behavior seen in Fig.~\ref{fig:b-and-bv_eff-const} remains generic as long as the $k^2$ terms decay more slowly than the $A(k)$ and $B(k)$ terms in Eqs.~\eqref{bias-result-g} and \eqref{bias-result-v}.
These $k^2$ terms come from spatial derivative in Boltzmann equation \eqref{vlasov-gal}, i.e., diffusion $\Delta$ and spatial and momentum derivative $\nabla\cdot \nabla_p$.
Thus, when the biased tracer undergoes an evolution in which diffusion remains effective, the corresponding bias parameters develop a characteristic $k^2$ dependence.
Meanwhile, when the $k$-independent coefficients $C_1$ and $C_2$ are non-zero, they shift the overall values of $b$ and $b_v$ upward or downward as seen at late times in Fig.~\ref{fig:b-and-bv_eff-const}.

Let us now consider a case that differs from Fig.~\ref{fig:b-and-bv_eff-const}, where the influence of the diffusion terms on the bias parameters decays over time.
In Fig.~\ref{fig:b-and-bv_C-k-linear}, the $k$-dependencies of $b$ and $b_v$ are specified by
\begin{align}
  &A(k)=0.004\times a_\mathrm{ini}^{3/2} 
  (k/H_0)\,, \qquad\qquad ~B(k)=0.012\times 
  a_\mathrm{ini} (k/H_0), 
  \label{Fig2 para1}
  \\
  &C_{gk}
  =C_{vk}
  =-10^{-5}(
  a/a_\mathrm{ini})^{-3/2},
  \qquad~~~C_g=C_v=0,
  \label{Fig2 para2}
\end{align}
where we again set $a_\mathrm{ini}=10^{-3}$.
With these conditions, $b$ and $b_v$ at $a=a_\mathrm{ini}$ are fixed as
\begin{align}
  &b(a_\mathrm{ini},k)=b_v(a_\mathrm{ini},k) = 1 +  0.04 (k/H_0) -10^{-3}(k/H_0)^2.
\end{align}
Thus, initially the density and velocity biases have linear and quadratic $k$-dependencies, which then decay as $a^{-1}$ or $a^{-3/2}$. 
Since we assume $C_{gk}(a)$ and $C_{vk}(a)$ decrease as $a^{-3/2}$, the initial $k$-dependence is eventually washed out in this case. This behavior is illustrated in Fig.~\ref{fig:b-and-bv_C-k-linear}.

It is interesting to compare our result \eqref{b sol} and \eqref{bv sol} with the peak bias model that predicts the bias parameters as~\cite{Desjacques:2009kt,Desjacques:2008jj,Matsubara:1999qq}
\begin{align}
  &b(a,k) = b_0 + b_1k^2,\qquad b_v(a,k) = 1 - R_v^2k^2\,.
  \label{peak-bias}
\end{align}
By setting $2C_1 + 3C_2=0$ which implies $C_v = 0$,
we reproduce this prediction with the coefficients
\begin{align}
  b_0 = 1 + \frac13 C_1,~~\qquad
  b_1=-\frac{5C_3+2C_4}{3H^2},~~\qquad
  R_v^2 = \frac{2C_3+2C_4}{3H^2}\,.
  \label{peak-compare}
\end{align}
This result is compatible with the literature \cite{Baldauf:2014fza}, which states that $b_v$ in Eq.~\eqref{peak-bias} is obtained through a modification of the Euler equation. 
However, BEA implies that only the $\nabla\cdot\nabla_p$ operator (the $C_4$ term) can modify the Euler equation without changing the continuity equation, whereas the $C_1$, $C_2$, and $C_3$ terms modify both equations, as seen in Eqs.~\eqref{continuity-eff} and \eqref{eff-euler}. In other words, one can modify only the Euler equation, as in Ref.~\cite{Baldauf:2014fza}, but in this framework $b_1$ and $R_v^2$ in Eq.~\eqref{peak-compare} are not independent. Thus, in a description where $b_1$ and $R_v^2$ are treated as independent parameters, the continuity equation should also be modified.

In the above argument, we imposed the relation 
$ 2C_1+3C_2=0$ 
such that the velocity bias vanishes on large scales.
This relation can be understood as a consequence of the generalized Galilean transformation associated with a long-wavelength growing mode.
According to the equivalence principle, local tracers cannot respond to a uniform gravitational acceleration or a common bulk flow generated by such a long-wavelength mode. Therefore, the collision term should not generate a physical source under the generalized Galilean transformation associated with a long-wavelength growing mode.

The generalized Galilean transformation is written as
\begin{align}
\label{eq:GGT}
    \bm x \rightarrow \tilde{\bm x}
    =
    \bm x+\bm d(t),
\end{align}
where the displacement $\bm d(t)$ is given by
\begin{align}
\bm d (t) = 
     \int^t \frac{{\bm u}_L (\tilde{t})}{a(\tilde t)} \, d\tilde{t} \,.
\end{align}
Here $\bm u_L$ denotes the bulk velocity for the long-wavelength growing mode, which satisfies the linear Euler equation,
\begin{align}
\label{eq:bulk-euler}
    \dot{\bm u}_L
    +
    H\bm u_L
    =
    -\frac1a\nabla\Phi_L .
\end{align}
Under the transformation~\eqref{eq:GGT}, the velocity and the gravitational potential transform as
\begin{align}
    \bm v
    &\rightarrow
    \tilde{\bm v}
    -
    \bm u_L ,
    \\
    -\frac1a\nabla\Phi
    &\rightarrow
    -\frac1a\nabla\tilde{\Phi}
    -
    \dot{\bm u}_L
    -
    H\bm u_L .
\end{align}
Keeping only the leading contribution in the large-scale limit ($k\rightarrow0$), where the $C_3$ and $C_4$ terms vanish, the collision term in the Euler equation,
\begin{align}
    {\cal F}HC_1\bm v
    -
    \frac{C_2}{a}\nabla\Phi ,
\end{align}
transforms as
\begin{align}
    {\cal F}HC_1\bm v
    -
    \frac{C_2}{a}\nabla\Phi
    \rightarrow
    {\cal F}HC_1\tilde{\bm v}
    -
    \frac{C_2}{a}\nabla\tilde{\Phi}
    -
    \left[
        {\cal F}HC_1\bm u_L
        +
        C_2
        \left(
            \dot{\bm u}_L
            +
            H\bm u_L
        \right)
    \right].
\end{align}
Using Eq.~\eqref{eq:bulk-euler} together with the growing-mode solution,
Eq.~\eqref{def-growth-factor}, and the Poisson equation, one obtains
\begin{align}
    \dot{\bm u}_L
    +
    H\bm u_L
    =
    -\frac1a\nabla\Phi_L
    =
    \frac32
    \frac{\Omega_m}{{\cal F}}
    H\bm u_L \,,
\end{align}
from which the transformation of the collision term can be written as
\begin{align}
    {\cal F}HC_1\bm v
    -
    \frac{C_2}{a}\nabla\Phi
    \rightarrow
    {\cal F}HC_1\tilde{\bm v}
    -
    \frac{C_2}{a}\nabla\tilde{\Phi}
    -
    \left[
        C_1
        +
        \frac32
        \frac{\Omega_m}{{\cal F}^2}
        C_2
    \right]
    {\cal F}H\bm u_L .
\end{align}
Requiring that the collision term is invariant under the transformation~\eqref{eq:GGT} associated with the long-wavelength growing mode, in other words, no zeroth-order velocity source is generated, we obtain
\begin{align}
    C_1
    +
    \frac32
    \frac{\Omega_m}{{\cal F}^2}
    C_2
    =
    0 .
\end{align}
For the matter-dominated (Einstein--de Sitter) Universe,
\begin{align}
    {\cal F}^2=\Omega_m=1 \,,
\end{align}
which reduces to
\begin{align}
    2C_1+3C_2=0 \qquad (C_v =0) \,.
\end{align}
Therefore no velocity bias is generated at zeroth order when we impose the generalized Galilean symmetry with long-wavelength growing mode.

\section{Implications for Observables \label{section:power-spectrum}}
In this section, we calculate the power spectrum based on the bias model~\eqref{bias-result-g} and \eqref{bias-result-v} in redshift-space distortions (RSD). 
Since every tracer has a peculiar velocity, observers cannot determine the exact position, but can only obtain the position in redshift space. In other words, the observed distribution depends on the velocity (e.g., see \cite{Heavens:1998es,Matsubara:2007wj}):
\begin{align}
\delta^s (\k) 
  = \delta_g (\k) - \frac{i\k\cdot\hat z}{aH}\v_{g}(\k)\cdot \hat z \,,
\end{align}
where $\hat z$ is a direction of the tracer.

It is common practice to apply the linear bias model which is valid for large scales as
\begin{align}
    &\delta_g (\k) = b_0 \delta(\k),
    \qquad 
    \v_g(\k) = \v(\k) = i\k\theta(\k)/k^2\cdot a\fh
      =i\k\delta(\k)/k^2\cdot a\fh \,,
\end{align}
where $b_0$ is a conventional bias parameter ($k-$independent) that is different from our result \eqref{bias-result-g}. 
Moreover, the velocity field is assumed to be irrotational, i.e., $\v(\k) \propto \k$.
In this regime, the density fluctuation is expressed as follows:
\begin{align}
     \delta^s = b_0\delta(\k) + \mathcal F \mu^2\delta(\k) \,,
\end{align}
where $\mu:= \k\cdot \hat z/k$ 
and the power spectrum of biased tracer $\ev{\delta^s(\k)\delta^s(\k')} :=(2\pi)^3\deltad(\k+\k')P^s(k)$ is
\begin{align}
    P^s(\k) = (b_0+\mathcal F\mu^2 )^{2}P(k) \,.
\end{align}
Here $\ev{\delta(\k)\delta(\k')} := (2\pi)^3\deltad(\k+\k')P(k)$ is the dark matter power spectrum.

This standard calculation is well established on large scales, whereas its predictions are known to deviate from simulation results toward smaller scales. To address this issue, the Effective Field Theory of Large-Scale Structure (EFTofLSS) has been developed in recent years, and the corresponding modification to the power spectrum is given by~\cite{Perko:2016puo}:
\begin{align}
    P_g^S
  & = \left[
    (b_0 + \mathcal F\mu^2)^2 
    -2\tilde c_0 k^2 -2\tilde c_2\mathcal F \mu^2k^2 -2 \tilde c_4 \mathcal F^2\mu^4 k^2
    -\tilde c \mathcal F^4 \mu^4 k^4 (b_0 + \mathcal F\mu^2)^2
    \right]P(k)\,.
    \label{Ps-EFT}
\end{align}
Here, the additional coefficients $\tilde c_0$, $\tilde c_2$, and $\tilde c_4$ are effective parameters that absorb small-scale nonlinear effects.
They correspond to the $\mu^0$, $\mu^2$, and $\mu^4$ angular structures, respectively, and model short-scale velocity dispersions and other redshift-space distortions. 
These $k^2$ terms are commonly included to incorporate contributions from higher-derivative bias \cite{Ivanov:2019pdj,Fujita:2016dne},
while the $k^4$-dependent term (i.e., the $\tilde c$ term in Eq.~\eqref{Ps-EFT}) was also added in \cite{Ivanov:2019pdj} 
in order to the fit to the data.

In contrast, applying our bias formula obtained in the previous section, \eqref{bias-result-g} and \eqref{bias-result-v}, we obtain
\begin{align}
  &\delta_g (k) = b(a,k) \delta(k),  \\[12pt]
  &  \v_g(k)\cdot \hat z 
      = b_v(a,k)\v(k)\cdot \hat z 
      = b_v(a,k)(iaH/k)(\mathcal F\mu\delta) ,\\[12pt]
  & 
  \begin{aligned}[b]
    \delta^s (\k) 
    &= 
    \delta_g (\k) - \frac{i\k\cdot \hat z}{aH}\v_{g}(\k)\cdot \hat z
    \\[6pt]
    &=
    \left[
      b(a,k) 
      +b_v(a,k) \mathcal F \mu^2
    \right]\delta(\k)\,, \\[6pt]
    & = 
      \left[
        (C_g+1) + (C_v+1) \mathcal F \mu^2
      \right]\delta(\k)
      +\left[
        C_{gk} + C_{vk} \mathcal F \mu^2
      \right]k^2\delta(\k),
  \end{aligned}
\end{align}
The power spectrum in the redshift space is:
\begin{align}
    P_g^s 
    &=
      \Bigl\{
        \left[(C_g+1) + (C_v+1)\mathcal F\mu^2)\right]^2 
        + 2(C_g+1) C_{gk} k^2
    \nonumber \\[8pt]
    &\qquad
        +2 \left[(C_g+1) C_{vk} + (C_v+1)C_{gk}\right]\mathcal F \mu^2 k^2 
        + 2(C_v+1)C_{vk}\mathcal F^2\mu^4 k^2
    \nonumber \\[8pt]
    &\qquad
        + (C_{gk} + C_{vk}\mathcal F\mu^2)^2k^4
      \Bigr\}P(k)\,.
    \label{Ps-BEA}
\end{align}

Now we compare the power spectrum obtained in the EFT~\eqref{Ps-EFT} and BEA~\eqref{Ps-BEA}.
\autoref{tab:compare} summarizes the terms of the two theories, organized according to their $k$-dependence.
\begin{table}[t]
\centering
\caption{
Comparison of the $k$-dependence between \eqref{Ps-BEA} and \eqref{Ps-EFT}.
}
\vspace{8pt}
\label{tab:compare}
\begin{tabular}{|l|c|c|}
\hline
 & \textbf{BEA} & \textbf{EFT} \\ \hline
$k^0$
& $\left[(C_g+1) + (C_v+1)\mathcal F\mu^2\right]^2$
& $(b_0 + \mathcal F\mu^2)^2$ \\ \hline
$k^2$
& $2(C_g+1)C_{gk}$
& $-2\tilde c_0$ \\ \hline
$\mathcal F\mu^2k^2$
& $2\left[(C_g+1)C_{vk} + (C_v+1)C_{gk}\right]$
& $-2\tilde c_2$ \\ \hline
$\mathcal F^2\mu^4k^2$
& $2(C_v+1)C_{vk}$
& $-2\tilde c_4$ \\ \hline
\end{tabular}
\end{table}
The first $k^0$ term can be compared with 
the EFT expression $(b_0+\mathcal F\mu^2)^2$. The corresponding one in the BEA result, with the large-scale velocity bias vanishing ($C_v=0$), matches the EFT expression upon identifying $b_0 = 1 + C_g$.
The second, third, and fourth terms, which have the $k^2, \mathcal F \mu^2 k^2$ and $\mathcal F^2\mu^4k^2$ structures, respectively, do not have fully independent coefficients in the present BEA. With the large-scale bias vanishing ($C_v=0$) and the relation $b_0 = 1 + C_g$ fixed by the $k^0$ term,  the three $k^2$ coefficients are determined by only two independent parameters, $C_{gk}$  and $C_{vk}$. 
This restriction originates from the fact that, in the present formulation,  we construct the effective collision term in real space and subsequently map the resulting density and velocity fields into redshift space. If, instead, the Boltzmann equation and its effective collision term were constructed directly in redshift space, additional line-of-sight-dependent operators allowed by symmetry, such as $(\hat z\cdot\nabla)^2 f$, would in general be present. Such operators can provide an additional independent contribution to the $\mu^4 k^2$ term, so that the number of independent $k^2$ coefficients can match that of the usual EFT description. Thus, the reduced number of independent coefficients in Eq.~\eqref{Ps-BEA} should not be regarded as an intrinsic reduction of the parameter freedom of the BEA. We leave a direct redshift-space formulation of the BEA for future work.

The fifth term scales as $k^4$. 
We emphasize that a $k^4$ dependence can in general arise from multiple independent contributions, including operators not included in the present truncation. 
Therefore, the $k^4$ dependence appearing in the present BEA expression should not be interpreted as a complete characterization at this order. 
Accordingly, this fit comparison is only a phenomenological comparison between the specific templates adopted in this work, and not a complete comparison of all possible $k^4$ contributions.

\section{Conclusion \label{section:conclusion}}

In this paper, we proposed an effective theory for biased tracers based on the Boltzmann equation approach and investigated its validity.

First, we confirmed that we can predict bias models, including peak theory, by 
introducing additional effective interaction terms into the Boltzmann equation. 
We showed that momentum derivative term with $\nabla\cdot \nabla_p$ only modifies the Euler equation, while the momentum Laplacian term with $\Delta_p$ affects neither the continuity nor the Euler equation.
We also applied spatial derivative as diffusion terms, 
which induce
$k$-dependencies in Fourier space. 

Second, we have calculated the power spectrum by applying 
the bias model in redshift space developed in Section~\ref{section:biased-boltzmann}  and compared it with those obtained in the EFT framework supplemented by linear bias without $k-$dependence.

In this study, we introduced very simple collision term that we can treat analytically. 
However, a variety of functions are generally allowed, and those possible terms should be further investigated.
Furthermore, although we limited our discussion only for the linear regime, higher order terms might give additional information.
Moreover, the mass dependence of the distribution function can, in principle, be incorporated into our formalism, although it was neglected in this paper.
Such a mass dependence may provide a link to the mass function.
These problems are put as future works.

\section*{Acknowledgements}
This work was partially supported by JSPS KAKENHI Grant Numbers 23K03424 (TF), 25K01004 (TT), MEXT KAKENHI 23H04515 (TT), 25H01543 (TT),   and Sasakawa Scientific Research Grant from
The Japan Science Society (SY).

\pagebreak

\bibliography{refs}

\end{document}